\documentclass[preprint,showpacs,preprintnumbers,amsmath,amssymb]{revtex4}
\usepackage{graphicx}
\usepackage{dcolumn}
\usepackage{bm}

\newcommand{\cL}{{\cal L}}
\newcommand{\cLg}{{\cal L}_{g}}
\newcommand{\p}{\partial}

\begin{document}

\title{Angular Momentum Conservation Law for Randall-Sundrum Models
\footnote{ This work was supported by the National Natural Science
Foundation of China.}}

\author{Yu-Xiao Liu}
\thanks{Corresponding author}
\email{liuyx@lzu.edu.cn} \affiliation{Institute of Theoretical
Physics, Lanzhou University, Lanzhou 730000, P. R. China}
\author{Yi-Shi Duan}
\affiliation{Institute of Theoretical Physics, Lanzhou University,
Lanzhou 730000, P. R. China}
\author{Li-Jie Zhang}
\affiliation{Department of Mathematics and Physics, Dalian Jiaotong University, Dalian
116028, P. R. China}

\begin{abstract}
In Randall-Sundrum models, by the use of general Noether theorem,
the covariant angular momentum conservation law is obtained with
the respect to the local Lorentz transformations. The angular
momentum current has also superpotential and is therefore
identically conserved. The space-like components $J_{ij}$ of the
angular momentum for Randall-Sundrum models are zero. But the
component $J_{04}$ is infinite.
\end{abstract}

\pacs{04.20.Cv, 04.20.Fy, 04.50.+h.\\
Keywords: Angular momentum, Randall-Sundrum models.} \maketitle


\section{Introduction}

Conservation laws of energy-momentum and angular momentum have
been of fundamental interest in gravitational physics
\cite{Penrose1982}. Using the vierbein representation of general
relativity, Duan ( one of the present authors ) {\em et al}
obtained a general covariant conservation law of energy-momentum
in 3+1 dimension which overcomes the difficulties of other
expressions \cite{Duan1963}. This conservation law gives the
correct quadrupole radiation formula of energy which is in good
agreement with the analysis of the gravitational damping for the
pulsar PSR1916-13 \cite{Duan1983}. Also, from the same point of
view, Duan and Feng \cite{DuanFeng1996} proposed a covariant
conservation law of angular momentum in four-dimensional Riemann
space-time which does not suffer from the flaws of the others
\cite{Fock1959,Ashtekar1982,Chevalier1990}.

Recently, there has been considerable activity in the study of
models that involve new extra dimensions. The possible existence
of such dimensions got strong motivation from theories that try to
incorporate gravity and  gauge interactions in a unique scheme, in
a reliable manner. The idea dates back to the 1920's,  to the
works of Kaluza and Klein \cite{Kaluza,Klein} who tried to unify
electromagnetism with Einstein gravity by assuming that the photon
originates from the fifth component of the metric. In the course
of the last several years, there has been active interest in the
brane world scenarios
\cite{Arkani-HamedPLB1998,AntoniadisPLB1998,CremadesNPB2002,KokorelisNPB2004,RS1,RS2,GarrigaPRL2000,GinddingsJHEP2000,LykkenJHEP2000,ShiromizuPRD2000}
and fermionic zero modes in Large dimensions
\cite{FrereJHEP2003,ourMPLA2005,ourMPLA2006}. The pioneering work
was done by Randall and Sundrum \cite{RS1,RS2}. In their works,
they present the so called Randall-Sundrum models
\cite{RS1,RS2,GogberashviliIJMPD2002,Arkani-HamedPRL2000} for
warped backgrounds, with compact or even infinite extra
dimensions. The RSI scenario provides a way to solve the hierarchy
problem, and the RSII scenario gives Newton's law of gravity on
the brane of positive tension embedded in an infinite extra
dimension.

As in the (3+1)-dimensional case, we should have some conservation
laws in order to understand high dimensional gravity well. In our
previous work \cite{LiuEnergy-Momentum2005}, we obtained the
general covariant conservation laws of energy-momentum in
(4+1)-dimensional Randall-Sundrum models. The purpose of the
present paper is to study the general covariant conservation law
of angular momentum for Randall-Sundrum models via the vierbein
representation. General relativity without vierbein is like a boat
without a jib---without these vital ingredients the going is slow
and progress inhibited. Consequently, vierbein has grown to be an
indispensable tool in many aspects of general relativity. More
important, it is relevant to the physical observability
\cite{FengHuang1997}. Based on the Einstein¡¯s observable time and
space interval, we take the local point of view that any
measurement in physics is performed in the local flat reference
system whose existence is guaranteed by the equivalence principle,
i.e. an observable object must carry the indices of the internal
space. Thus, we draw the support from vierbein not only for
mathematical reasons, but also because of physical measurement
consideration.

This paper is arranged as follows. In section \ref{section2}, we
give a general description of the scheme for establishing general
covariant conservation laws in general relativity. In section
\ref{section3}, we first give a simple review of the
Randall-Sundrum models. Then use local $SO(1,4)$ transformations
and the scheme in section \ref{section2} to obtain a covariant
conservation law of angular momentum for Randall-Sundrum models.
Finally, we calculate the angular momentum of the bulk for
Randall-Sundrum solution by superpotential. Section
\ref{sectionDiscussions} is devoted to some remarks and
discussions.

\section{General conservation laws in general relativity}\label{section2}
The conservation law is one of the important problems in
gravitational theory. It is due to the invariance of the action
corresponding to some transformations. In order to study the
covariant angular momentum conservation law of more complicated
systems, it is necessary to discuss conservation laws by Noether
theorem in the general case
\cite{Duan1963,Duan1987,Duan1988,Feng1999,DuanFengAPS1995,Cho1995}.
Suppose that the space-time manifold $\cal M$ is of dimension
$n=1+d$ and the Lagrangian density is in the first order
formalism, i.e.
\begin{equation} \label{action}
I=\int_{\cal M} d^{n}x \cL(\phi^{A}, \p_{\mu}\phi^{A}),
\end{equation}
where $\phi^{A}$ denotes the general fields. If the action is
invariant under the infinitesimal transformations
\begin{eqnarray}
x^{\prime\mu} &=& x^{\mu}+\delta x^{\mu}, \label{transformation1}\\
\phi^{\prime A}(x^{\prime}) &=&
\phi^{A}(x)+\delta\phi^{A}(x),\label{transformation2}
\end{eqnarray}
and $\delta\phi^{A}$ vanishes on the boundary of $\cal M$,
$\partial \cal M$, then following relation holds
\cite{Duan1963,Duan1988,Feng1995}
\begin{equation} \label{NoetherTheorem}
\p_{\mu}(\cL\delta x^{\mu}+\frac{\p\cL}{\p\p_{\mu}\phi^{A}}
\delta_{0}\phi^{A} )+[\cL]_{\phi^{A}}\delta_{0}\phi^{A}=0,
\end{equation}
where $[\cL]_{\phi^{A}}$ is
\begin{equation}
[\cL]_{\phi^{A}}=\frac{\p\cL}{\p\phi^{A}}-\p_{\mu}
\frac{\p\cL}{\p\p_{\mu}\phi^{A}},
\end{equation}
and $\delta_{0}\phi^{A}$ is the Lie derivative of $\phi^{A}$
\begin{equation}\label{LieDerivative}
\delta_{0}\phi^{A}=\phi^{\prime A}(x)-\phi^{A}(x)=
\delta\phi^{A}(x)-\p_{\mu} \phi^{A}\delta x^{\mu}.
\end{equation}

If $\cL$ is the total Lagrangian density of the system, the field
equation of $\phi^{A}$ is just $[\cL]_{\phi^{A}}=0$. Hence from
Eq. (\ref{NoetherTheorem}), we can obtain the conservation
equation corresponding to transformations (\ref{transformation1})
and (\ref{transformation2})
\begin{equation} \label{ConservationEq}
\p_{\mu}(\cL\delta x^{\mu}+\frac{\p\cL}{\p\p_{\mu}\phi^{A}}
\delta_{0} \phi^{A})=0.
\end{equation}
This is just the conservation law in general case. It is important
to recognize that if $\cL$ is not the total Lagrangian density of
the system, e.g. the gravitational part $\cL_{g}$, then so long as
the action of $\cL_{g}$ remains invariant under transformations
(\ref{transformation1}) and (\ref{transformation2}), Eq.
(\ref{NoetherTheorem}) is still valid yet Eq.
(\ref{ConservationEq}) is no longer admissible because of
$[\cL_{g}]_{\phi^{A}}\not=0$.

In a gravitational theory with the vierbein as elementary fields,
we can separate $\phi^{A}$ as $\phi^{A}=(e^{\mu}_{a}, \psi^{B})$,
where $e^{\mu}_{a}$ is the vierbein field and $\psi^{B}$ is an
arbitrary tensor under general coordinate transformations. When
$\psi^{B}$ is $\psi^{\mu_{1}\mu_{2} \cdots \mu_{k}}$, we can
always scalarize it by
\begin{equation}
\psi^{a_{1}a_{2} \cdots
a_{k}}=e^{a_{1}}_{\mu_{1}}e^{a_{2}}_{\mu_{2}}\cdots
e^{a_{k}}_{\mu_{k}} \psi^{\mu_{1}\mu_{2} \cdots \mu_{k}},
\nonumber
\end{equation}
so we can take $\psi^{B}$ as a scalar field under general
coordinate transformations. In later discussion we can simplify
the equations by such a choice.

\section{Covariant angular momentum conservation law for
Randall-Sundrum models}\label{section3}

In this section, we first give a brief introduction of the
Randall-Sundrum background. Then, with these foundations above, we
use local $SO(1,4)$ transformations to obtain the covariant
angular momentum conservation law for Randall-Sundrum models.

\subsection{Randall-Sundrum background}\label{subsection3A}

Let us consider the following setup. A five dimensional spacetime
with an orbifolded fifth dimension of radius $r$ and coordinate
$y$ which takes values in the interval $[0,\pi r]$. Consider two
branes at the fixed (end) points $y=0,\pi r$; with tensions $\tau$
and $-\tau$ respectively. The brane at $y=0$ ($y=\pi r$) is
usually called the hidden (visible) or Planck (SM) brane. We will
also assign to  the bulk  a negative cosmological constant
$-\Lambda$. Here we shall assume that all parameters are of the
order the Planck scale.

The classical action describing  the above setup is given by
\begin{equation} \label{ActionOfRS}
S= S_{g} + S_{h} + S_{v},
\end{equation}
here
\begin{equation}
S_{g} = \int \! d^4x \, dy \sqrt{g} \left( \frac{1}{2k_\ast^2} R
+\Lambda\right)
\end{equation}
gives the bulk contribution, whereas the visible and hidden brane
parts are given by
\begin{equation}
S_{v,h}= \pm~\tau~\int\! d^4x \sqrt{-g_{v,h}}~,
\end{equation}
where $g_{v,h}$ stands for the induced metric at the visible and
hidden branes, respectively. And $2k_\ast^2=8\pi
G_\ast=M_\ast^{-3}$. Five dimensional  Einstein equations for the
given action are
\begin{eqnarray} \label{eers}
 G_{\mu\nu} &=&-\; k_\ast^2\Lambda\,g_{\mu\nu} +
   k_\ast^2\tau\, \sqrt{\frac{-g_{h}}{g}}
    \delta_{\mu}^{\bar{\mu}} \delta_{\nu}^{\bar{\nu}} g_{\bar{\mu}\bar{\nu}}\delta(y) \nonumber \\
    && -\; k_\ast^2\tau\, \sqrt{\frac{-g_{v}}{g}}
    \delta_{\mu}^{\bar{\mu}} \delta_{\nu}^{\bar{\nu}} g_{\bar{\mu}\bar{\nu}}\delta(y-\pi r),
\end{eqnarray}
 where the Einstein tensor $G_{\mu\nu} = R_{\mu\nu} - \frac{1}{2} g_{\mu\nu}
 R$ as usual, Greek indices without bar $\mu, \nu = 0, \cdots, 4$
 and the others with bar $\bar{\mu}, \bar{\nu} = 0, \cdots, 3$.
 The solution that gives a flat induced metric on the branes is
\begin{equation}\label{rsmetric}
ds^2 = g_{\mu\nu}dx^\mu dx^\nu = e^{-2k
|y|}\eta_{\bar{\mu}\bar{\nu}}dx^{\bar{\mu}} dx^{\bar{\nu}} -
dy^2~,
\end{equation}
in which $x^{\bar{\mu}}$ are coordinates for the familiar four
dimensions, $k$ is a scale of order of the Planck length
\begin{equation} \label{rsmu}
k^2 =\frac{k_\ast^2\Lambda}{6}= \frac{\Lambda}{6 M_\ast^3}~,
\;\;\; \Lambda = \frac{\tau^2}{6 M_\ast^3}~.
\end{equation}
The effective Planck scale in the theory is given by
\begin{equation}\label{kMpM*}
 M_{P}^2 = \frac{M_\ast^3}{k}\left(1- e^{-2k\pi r}\right).
\end{equation}
Notice that for large $r$, the exponential piece becomes
negligible, and above expression has the familiar form given in
ADD models
\cite{Arkani-HamedPLB1998,AntoniadisPLB1998,CremadesNPB2002,KokorelisNPB2004}
for one extra dimension of (effective) size $R_{ADD}=1/k$:
\begin{equation}
 M_{P}^2 = M_\ast^{2+n} R_{ADD}^n.
 \label{rsmp}
\end{equation}

\subsection{Covariant conservation law of angular momentum}\label{subsection3B}

It is well known that in deriving the general covariant conservation law of
energy-momentum in general relativity, the general displacement transformations, which is
a generalization of the displacement transformations in the Minkowski space-time, was
used \cite{Feng1995}. In the local Lorentz reference frame, the general displacement
transformations take the same form as that in the Minkowski space-time. This implies that
general covariant conservation laws are corresponding to the invariant of the action
under local transformations. We may conjecture that since the conservation law for
angular momentum in special relativity corresponds to the invariance of the action under
the Lorentz transformations, the general covariant conservation law of angular momentum
in general relativity may be obtained by means of the local Lorentz invariance.

The Lagrangian density for Randall-Sundrum background can be
written as
\begin{equation}
\cL=\cLg+\cL_{h}+\cL_{v}+\cL_{m},
\end{equation}
where $\cL_{g}$ is the Lagrangian density of gravity which gives
the bulk contribution, whereas $\cL_{h}$ and $\cL_{v}$ to the
hidden and visible brane parts, respectively, and $\cL_{m}$
denotes the matter part
\begin{eqnarray}
\cLg&=&\sqrt{g} \left( \frac{1}{2k_\ast^2} R +\Lambda\right),\\
\cL_{h}&=& \tau \sqrt{-g_{h}} \; \delta(y-\pi r), \\
\cL_{v}&=& -\tau \sqrt{-g_{v}} \; \delta(y),\\
\cL_{m}&=&\cL_{m}(\phi^{A},D_{\mu}\phi^{A})\label{Lm}.
\end{eqnarray}
The matter fields $\phi^{A}$ belong to some representation of
$SO(1,4)$ whose generators are $I_{ab}(a,b=0,1,2,3,4)$ and
$I_{ab}=-I_{ba}$. $D_{\mu}$ is the local $SO(1,4)$ gauge covariant
derivative operator of $\phi^{A}$
\begin{equation}
D_{\mu}\phi^{A}=\partial_{\mu}\phi^{A}-\frac{1}{2}\omega_{\mu
ab}(I^{ab})^{A}_{\;\; B}\phi^{B}.
\end{equation}
Under the local $SO(1,4)$ gauge transformations
\begin{equation}\label{TransformOfVierbein}
e^{a}_{\mu}(x) \rightarrow e'^{a}_{\mu}(x)=\Lambda^{a}_{\;\;b}(x)
e^{b}_{\mu}(x),  \;\;\;\;
\eta_{ab}\Lambda^{a}_{\;\;c}\Lambda^{b}_{\;\;d}=\eta_{cd}.
\end{equation}
$\phi^{A}$ transforms as \cite{KastorMartine1989}
\begin{equation}
\phi^{A}(x) \rightarrow \phi '^{A}(x)=D(\Lambda(x))^{A}_{\;\;B}
\phi^{B}(x).
\end{equation}
The infinitesimal $SO(1,4)$ rotations can be linearized as follows
\begin{equation}
\Lambda^{a}_{\;\;b}(x)=\delta^{a}_{\;\;b}+\alpha^{a}_{\;\;b}(x),
\;\;\;\; \alpha_{ab}=-\alpha_{ba}.
\end{equation}
$D(\Lambda)$ can be linearized as
\begin{equation}
[D(\Lambda)]^{A}_{\;\;B}=\delta^{A}_{\;\;B}+\frac{1}{2}(I_{ab})^{A}_{\;\;B}\alpha^{ab}.
\end{equation}
Thus the variation of $\phi^{A}$ is
\begin{equation}
\delta\phi^{A}=\phi
'^{A}-\phi^{A}=\frac{1}{2}(I_{ab})^{A}_{\;\;B}\alpha^{ab}\phi^{B}.
\end{equation}
As in the (3+1)-dimensional case, we have the following
decomposition
\begin{equation}
\cL=\cL_{\omega}+\cL_{\Delta}+\cL_{b}+\cL_{m},
\end{equation}
where
\begin{eqnarray}
\cL_{\omega}&=&\frac{1}{2k^{2}_{*}}(\omega_a
\omega^a-\omega_{abc}\omega^{cba})\sqrt{g},\label{Lomega}\\
\cL_{\Delta}&=& - \frac{1}{2k^{2}_{*}}\;\p_{\mu}(
 \sqrt{g}\;e^{a\mu}\p_{\nu}e^{\nu}_{a}
-\sqrt{g}\;e^{\nu}_{a}\p_{\nu}e^{a\mu}),\label{LDelta}\\
\cL_{b}\;&=& \Lambda \sqrt{g} - \tau \sqrt{-g_{v}} \; \delta(y) +
\tau \sqrt{-g_{h}} \; \delta(y-\pi r),\label{Ls}\\
\omega_{abc}&=& \frac{1}{2} (\Omega_{abc}-\Omega_{bca}+\Omega_{cab}),\nonumber\\
\Omega_{abc}&=& e^{\mu}_{a} e^{\nu}_{b}(\p_{\mu} e_{c\nu}-\p_{\nu} e_{c\mu}),\nonumber\\
\omega_{a}\;&=& \eta^{bc} \omega_{bac}=\omega^{c}_{\;\;ac}.
\nonumber
\end{eqnarray}

We choose vierbein $e^{a}_{\mu}$ and the matter field $\phi^{A}$
as independent variables. Since the coordinates $x^{\mu}$ do not
transform under the local Lorentz transformations, $\delta
x^{\mu}=0$, from Eq. (\ref{LieDerivative}), it can be proved that
in this case, $\delta_{0} \rightarrow \delta$. It is required that
$\cL_{m}$ is invariant under (\ref{TransformOfVierbein}) and
$\cL_{\omega}$, $\cL_{\Delta}$ and $\cL_{s}$ are invariant
obviously. So under the local Lorentz transformations
(\ref{TransformOfVierbein}) $\cL$ is invariant. In the light of
the discussion in section II, we would like to the relation
\begin{eqnarray} \label{Relation1}
  \frac{\p}{\p x^{\mu}}
              \left(  \frac{\p\cL}{\p\p_{\mu}e^{\nu}_{a}} \; {\delta}e^{\nu}_{a}
                     +\frac{\p\cL}{\p\p_{\mu}\phi^{A}} \; {\delta}\phi^{A}
              \right)
  + [\cL]_{e^{\nu}_{a}}\delta e^{\nu}_{a}+[\cL]_{\phi^{A}}{\delta}\phi^{A} = 0,
\end{eqnarray}
where $[\cL]_{e^{\nu}_{a}}$ and $[\cL]_{\phi^{A}}$ are the Euler
expressions defined as
\begin{eqnarray}
[ \cL ]_{e^{\nu}_{a}}&=&\frac{\p\cL}{\p e^{\nu}_{a}}-\p_{\mu}
\frac{\p\cL}{\p\p_{\mu}e^{\nu}_{a}}, \nonumber \\
\;[ \cal L ] _{\phi^{A}}&=&\frac{\p\cL}{\p\phi^{A}}-\p_{\mu}
\frac{\p\cL}{\p\p_{\mu}\phi^{A}}. \nonumber
\end{eqnarray}
Using the Einstein equation $[\cL]_{e^{\nu}_{a}}=0$ and the
equation of motion of matter $[\cL]_{\phi^{A}}=0$, we get
following equation by (\ref{Relation1})
\begin{eqnarray} \label{Relation2}
\frac{\p}{\p x^{\mu}}
              \left(\frac{\p\cL_{\omega}}{\p\p_{\mu}e^{\nu}_{a}}\;{\delta}e^{\nu}_{a}
                    +\frac{\p\cL_{\Delta}}{\p\p_{\mu}e^{\nu}_{a}}\;{\delta}e^{\nu}_{a}
                    +\frac{\p\cL_{b}}{\p\p_{\mu}e^{\nu}_{a}}\;{\delta}e^{\nu}_{a}
              \right) \nonumber\\
+ \frac{\p}{\p x^{\mu}}
              \left(  \frac{\p\cL_{m}}{\p\p_{\mu}e^{\nu}_{a}} \; {\delta}e^{\nu}_{a}
                     +\frac{\p\cL_{m}}{\p\p_{\mu}\phi^{A}} \; {\delta}\phi^{A}
              \right)
  = 0,
\end{eqnarray}
where we have used the fact that only $\cL_{m}$ contains the
matter field $\phi^{A}$.

We introduce $j^{\mu}_{ab}$
\begin{eqnarray} \label{jmuab}
  \sqrt{g}j^{\mu}_{ab} =
              \left(\frac{\p\cL_{\omega}}{\p\p_{\mu}e^{a\nu}}\;e^{\nu}_{b}
              + \frac{\p\cL_{b}}{\p\p_{\mu}e^{a\nu}}\;e^{\nu}_{b}
              + \frac{\p\cL_{m}}{\p\p_{\mu}e^{a\nu}} \; e^{\nu}_{b}
              + \frac{1}{2}\frac{\p\cL_{m}}{\p\p_{\mu}\phi^{A}} \; (I_{ab})^{A}_{\;\;B}\phi^{B}
              \right),
\end{eqnarray}
then (\ref{Relation2}) can be rewritten as
\begin{eqnarray} \label{Relation3}
 \p_{\mu}(\sqrt{g}j^{\mu}_{ab}\alpha^{ab})
 +\p_{\mu}\left(\frac{\p\cL_{\Delta}}{\p\p_{\mu}e^{\nu}_{a}}\;e^{b\nu}\alpha_{ab}\right)=0.
\end{eqnarray}
From (\ref{LDelta}) one can get easily that
\begin{eqnarray}
\frac{\p\cL_{\Delta}}{\p\p_{\lambda}e^{\nu}_{a}}\;e^{b\nu}\alpha_{ab}
 =-\frac{1}{2k^{2}_{*}}\alpha^{ab}\p_{\mu}(\sqrt{g}\;V^{\mu\lambda}_{ab}),
\end{eqnarray}
where
\begin{eqnarray}
V^{\mu\lambda}_{ab}=e^{\mu}_{a}e^{\lambda}_{b}-e^{\mu}_{b}e^{\lambda}_{a}.\label{V}
\end{eqnarray}
Substituting (\ref{V}) into (\ref{Relation3}), we obtain
\begin{eqnarray} \label{Relation4}
 \p_{\mu}(\sqrt{g}j^{\mu}_{ab})\alpha^{ab}
 +\left[ \sqrt{g}j^{\mu}_{ab}
   -\frac{1}{2k^{2}_{*}}\p_{\nu}(\sqrt{g}\;V^{\nu\mu}_{ab})
  \right]\p_{\mu}\alpha^{ab}
 =0.
\end{eqnarray}
Since $\alpha^{ab}$ and $\p_{\mu}\alpha^{ab}$ are independent of
each other, we must have
\begin{eqnarray}
& \p_{\mu}(\sqrt{g}j^{\mu}_{ab})=0, \label{pj}\\
& j^{\mu}_{ab}=\frac{1}{2k^{2}_{*}}  \bigtriangledown_{\nu}V^{\nu\mu}_{ab}.\label{j}
\end{eqnarray}
From (\ref{pj}) and (\ref{j}), it can be concluded that
$j_{ab}^\mu $ is conserved identically. As usual, we call
$V_{ab}^{\nu \mu }$ superpotential. Since the current $j_{ab}^\mu
$ is derived from the local Lorentz invariance of the total
Lagrangian, it can be interpreted as the total angular momentum
tensor density of the gravity-matter system, and it
contains the spin density of the matter field: $(\partial {\cal L}%
_m)/(\partial \partial _\mu e_a^\nu ).$ From the above discussion, we see that not only
the current $j_{ab}^\mu $ but also the superpotential $V_{ab}^{\mu \nu }$ does not has
any terms relevant to the visible and hidden branes, both of them are only determined by
the vierbein.

For a globally hyperbolic manifold $\cal{M}$, there exist  a
series of Cauchy surfaces $\Sigma _t$ foliating $\cal{M}$. We
choose a submanifold $D$ of $\cal{M}$ joining any two Cauchy
surfaces $\Sigma _{t_1}$ and $\Sigma _{t_2}$ so the boundary
$\partial D$ of $D$ consists of six parts $\Sigma _{t_1},\;\Sigma
_{t_2},\;B_{v},\;B_{h},\;A_{1}$ and $A_{2}$, in which $\Sigma
_{t_1}$ and $\Sigma _{t_2}$ are Cauchy surfaces, $B_{v}$ and
$B_{h}$ are the visible and hidden branes, respectively, $A_{1}$
and $A_{2}$ are at the spatial infinity of the two branes.
For the solution (\ref{rsmetric}), we can obtain the following
vierbein
\begin{equation} \label{vierbein}
e^{a}_{\mu} = (e^{-k \mid y
\mid}\delta^{\bar{a}}_{\mu},\delta^{4}_{\mu}). \;\;\;
(\bar{a}=0,1,2,3)
\end{equation}
So, the following relations are tenable
\begin{eqnarray}
&(\p_\mu e_{a\nu}-\p_\nu e_{a\mu})|_{B_{v,h}}=0,
\label{Condition1}\\
&(\p_\mu e_{a\nu}-\p_\nu e_{a\mu})|_{A_{1}}=(\p_\mu e_{a\nu}-\p_\nu
e_{a\mu})|_{A_{2}}.\label{Condition1}
\end{eqnarray}
Since
$$
\sqrt{g}\;V_{ab}^{\mu \nu }=\frac 12\epsilon ^{\mu \nu \lambda
\rho }\epsilon _{abcd}e_{c\lambda}e_{d\rho},
$$
we have
\begin{eqnarray}
&\partial _\lambda (\sqrt{g}\;V_{ab}^{\lambda \mu
})|_{B_{v,h}}=0,\\
&\partial_\lambda(\sqrt{g}\;V_{ab}^{\lambda
\mu})|_{A_{1}}-\partial_\lambda(\sqrt{g}\;V_{ab}^{\lambda \mu})|_{A_{2}}=0.
\end{eqnarray}
Thus, by the use of the following identity
\begin{equation}
\int_{D}(\triangledown_{\mu}j^{\mu}_{ab}) \sqrt{g}\;dx^5 =0,
\end{equation}
we can get the total conservative angular momentum from (\ref{pj})
and (\ref{j})
\begin{equation}\label{Jab1}
J_{ab}=\int_{\Sigma _t}j_{ab}^\mu \sqrt{g}\;d\Sigma _\mu =\frac{1}{2k^{2}_{*}}%
\int_{\partial \Sigma _t}\sqrt{g}\;V_{ab}^{\mu \nu }d\sigma _{\mu
\nu },
\end{equation}
where $\sqrt{g}\;d\Sigma _\mu $ is the covariant surface element
of $\Sigma _t $, $d\Sigma _\mu =\frac{1}{4!}\epsilon _{\mu \nu
\lambda \rho \beta}dx^\nu \wedge dx^\lambda \wedge dx^\rho \wedge
dx^\beta$, $d\sigma _{\mu \nu }=\frac{1}{3!}\epsilon _{\mu \nu
\lambda \rho \beta}dx^\lambda \wedge dx^\rho \wedge dx^\beta.$
Because $dx^{0}=dt=0$ on the Cauchy surface $\Sigma _{t}$, the
expression (\ref{Jab1}) can be rewritten as following
\begin{equation}\label{Jab2}
J_{ab}=\int_{\Sigma _t}j_{ab}^{0} \sqrt{g}\;dx^{1}dx^{2}dx^{3}dy.
\end{equation}
The calculation result of $J_{ab}$ is
\begin{eqnarray}
J_{04}&=&\frac{1}{2k^{2}_{*}}V_{3D}(1-e^{-k\pi r}),\label{JabResult1}\\
J_{0\bar{i}}&=&0, \;\;\; (\bar{i}=1,2,3)\label{JabResult2}\\
J_{ij}&=&0, \;\;\; (i,j=1,2,3,4) \label{JabResult2}
\end{eqnarray}
where $V_{3D}$ stands for the volume of usual three dimensional
space
\begin{equation}
V_{3D}=\int d^3 x.
\end{equation}
So the space-like components the angular momentum are zero. But the non-space-like
component $J_{04}$ of it is infinite, this is caused by the gravity on the warped extra
dimension. When the radius $r$ of the extra dimension is taken the limit $r \rightarrow
\infty$, $J_{04}=\frac{1}{2k^{2}_{*}}V_{3D}$.

\section{Discussions}\label{sectionDiscussions}
To summarize, by the use of general Noether theorem, we have obtained the conservation
law of angular momentum for the Randall-Sundrum models with the respect to local
$SO(1,4)$ transformations.  This conservation law is a covariant theory with respect to
the generalized coordinate transformations, but the angular momentum tensor is not
covariant under the local Lorentz transformation which, due to the equivalent principle,
is reasonable to require.

The angular momentum current has also superpotential and is therefore identically
conserved. The conservative angular momentum current and the corresponding superpotential
for the Randall-Sundrum models are the same with those in (3+1)- and (2+1)-dimensional
Einstein theories, the Lagrangian density $\cL_{h,v}$ corresponding to the hidden and
visible brane parts do not play a role in the conservation law. Both angular momentum
current and the superpotential are determined only by vierbein field.

\section*{Acknowledgement}
It is a pleasure to thank Dr Liming Cao, Yongqiang Wang, Zhenhua
Zhao and Zhenbin Cao for interesting discussions.

\end{document}